\documentclass[preprint]{revtex4}
\usepackage[latin1]{inputenc}
\usepackage{graphicx,epstopdf}
\usepackage{subfigure}
\usepackage{epsfig}
\usepackage{bigints}
\usepackage[usenames]{color}

\begin{document}

\title{Compact stars with strongly coupled quark matter in a strong magnetic field}

\author{D. A. Foga\c{c}a, S. M. Sanches Jr., T. F. Motta and F. S. Navarra}

\address{Instituto de F\'isica, Universidade de S\~ao Paulo,
Rua do Mat\~ao Travessa R, 187, 05508-090 S\~ao Paulo, SP, Brazil}

\begin{abstract}

Some time ago we have derived from the QCD Lagrangian an equation of state (EOS) for the cold quark matter, which can be considered an
improved version of the MIT bag model EOS. Compared to the latter, our equation of state reaches higher values of the pressure at
comparable baryon densities. This feature is due to perturbative corrections and also to non-perturbative effects. Later we applied
this EOS to the study of compact stars, discussing the absolute stability of quark matter and computing the mass-radius relation
for self-bound (strange) stars. We found maximum masses of the sequences with more than two solar masses, in agreement with the recent
experimental observations. In the present work we include the magnetic field in the equation of state and study how it changes the stability
conditions and the mass-radius curves.

\end{abstract}

\maketitle

\section{Introduction}

In the theory of compact stars \cite{hebel,emmi,kojo,fukojo,drago,fkv16,kfbv,fkv14}  there are still several  key unanswered questions \cite{emmi}.
One of them is: ``are there quark stars  ?''  This question has been around for decades and it has received a
renewed attention after the appearance of new measurements of masses of astrophysical compact objects \cite{demorest,anton,vanker}.
These measurements suggest that stellar objetcs may have large masses, such as, for example,
the pulsar PSR J1614-2230, with $(1.97 \pm 0.04) \, M_{\bigodot}$ \cite{demorest} or the pulsar PSR J0348+0432, with
$(2.01 \pm 0.04) \, M_{\bigodot}$ \cite{anton} and perhaps the black widow pulsar PSR B1957+20, with a possible mass around
$(2.4 \pm 0.12) \, M_{\bigodot}$ \cite{vanker}. In principle larger masses imply larger baryon densities in the core of the stars and we
expect very dense hadronic matter to be in a quark gluon plasma (QGP) phase. On the other hand, from the theoretical point of view, most of the proposed
equations of state for this cold QGP are too soft to be able to support such large masses.

The answer to the question above depends on the details of the equation of state of cold quark matter. According to most models, deconfined quark matter should
be formed at baryon densities in the range $\rho_B = 2 \rho_0 - 5 \rho_0$, where $\rho_0$ is the ordinary nuclear matter baryon density. Since at low
temperatures and high baryon densities we can not rely on lattice QCD calculations,  the  quark matter equations of state must be derived from
models.  Many of them are based on the MIT bag model \cite{mit}  or on the Nambu-Jona-Lasinio (NJL) model \cite{nambu}.  In these models the
gluon degrees of freedom do not appear explicitly. In the bag model they are contained in the bag constant and in the NJL they are integrated out giving
origin to the four-quark terms. In more recent version of the NJL model \cite{kojo} a bag-like term was introduced to represent the contribution of gluons
to the pressure and energy density. At very high baryon densities there are constraints derived from perturbative QCD calculations  \cite{fkv16,kfbv,fkv14,pqcd}.
In Refs.  \cite{we11,we111} we have developed a quark-gluon EOS, which was applied to
the calculation of the structure of compact quark stars  in  Ref. \cite{we12}.  Stars as heavy as $2 \, M_{\bigodot}$ were found.

One important ingredient in the stellar structure calculation is the magnetic field. In magnetars, the strength of this field can reach values as large as
$10^{18} \, G$. In theoretical calculations, magnetic field effects in astrophysical compact objects have been well studied
\cite{magalera,dpm3}.

In this work we extend the equation of state derived in Ref. \cite{we11} to the case where we have strong magnetic fields  and check whether it is
still able to support massive stars. Some steps along this direction have already been taken in Ref. \cite{we15}.

\section{Equation of state}

The equation of state derived in \cite{we11} is based on a few assumptions.  First we assume, as in the case of the hot QGP observed in heavy  ion
collisions, that the quarks and gluons in the cold QGP are deconfined but ``strongly interacting'', forming a strongly interacting QGP (sQGP). This
means that the coupling is not small and also that there are remaining non-perturbative interactions and  gluon condensates.
Of course, at very large densities (in the same way as at very  high temperatures) the sQGP evolves to an ideal gas of non-interacting particles in a
trivial vacuum. We split the gluon field  into two components  $G^{a\mu}={A}^{a\mu}+{\alpha}^{a\mu}$, where ${A}^{a\mu}$ (``soft'' gluons) and
${\alpha}^{a\mu}$ (``hard''gluons) are the components of the field associated with low and high momentum respectively. The expectation values of
${A}^{a\mu}{A}^{a}_{\mu}$ and ${A}^{a\mu}{A}^{a}_{\mu} {A}^{b\nu}{A}^{b}_{\nu}$ are non-vanishing in a non-trivial vacuum and from them we  obtain
an effective gluon mass ($m_{G}$)
and also a contribution (${\mathcal{B}}_{QCD}$) to the energy and to the pressure  of the system similar to the one of the MIT bag model. Since the number
of quarks is very large and their coupling to the gluons is not small, the high momentum levels of the gluon field will have large occupation
numbers and hence the ${\alpha}^{a\mu}$ component of the field can be approximated by a classical field. This is the same mean field approximation
very often applied to  models of nuclear matter, such as the Walecka model.

In the next subsection we review the main formulas. For the details of the
derivation we refer the reader to Ref. \cite{we11}.

\subsection{Effective Lagrangian}

Let us consider a system of deconfined quarks and gluons in a non-trivial vacuum immersed in an homogeneous magnetic field oriented along  the Cartesian $z$ direction
(we employ natural units $\hbar=c=k_{B}=1$ and  metric  given by $g_{\mu\nu}=\textrm{diag}(+,-,-,-)$) :
\begin{equation}
\vec{B}=B \hat{z} \hspace{1.2cm} \textrm{and}  \hspace{1.2cm} A_{\mu}=(0,yB,0,0)
\label{magfield}
\end{equation}
The Lagrangian is given by:
\begin{equation}
{\mathcal{L}}={\mathcal{L}}_{QCD} \, + \, {\mathcal{L}}_{QED}
\label{prelagra}
\end{equation}
where ${\mathcal{L}}_{QCD}$, ${\mathcal{L}}_{QED}$ refer to the  quarks  and electrons  which interact with the external magnetic field.
The electrons are necessary to ensure the charge neutrality of the  star, which will be enforced as in \cite{we12}. The Lagrangian (\ref{prelagra}) can be written as:
$$
{\mathcal{L}}
=-{\frac{1}{4}}F^{a}_{\mu\nu}F^{a\mu\nu}
+\sum_{f=u}^{d,s}\bar{\psi}_{i}^{f}\Big[i\gamma^{\mu}(\delta_{ij}\partial_{\mu}
+i\delta_{ij}Q_{f}A_{\mu}
-igT^{a}_{ij}G_{\mu}^{a})-\delta_{ij}m_{f}\Big]\psi_{j}^{f}
$$
\begin{equation}
+\bar{\psi}_{i}^{e}\Big[i\gamma^{\mu}(\delta_{ij}\partial_{\mu}+i\delta_{ij}Q_{e}A_{\mu}
)-\delta_{ij}m_{e}\Big]\psi_{j}^{e}
-{\frac{1}{16\pi}}F_{\mu\nu}F^{\mu\nu}
\label{lqcduagain}
\end{equation}
where the first and second lines represent the QCD and QED parts respectively.
The summation in $f$ runs over the quark flavors: up ($u$), down ($d$) and strange ($s$), which have the following masses: $m_{u}=5 \, MeV$,
$m_{d}=7 \, MeV$  and $m_{s}=150 \, MeV$. The electron mass is $m_{e}=0.5 \, MeV$.
The respective charges are : $Q_{u}= 2 \, Q_{e}/3$, $Q_{d}= - \, Q_{e}/3$  and $Q_{s}= - \, Q_{e}/3$, where $Q_{e}$ is the absolute value of the electron charge.
$T^{a}$ are the SU(3) generators, $f^{abc}$ are the SU(3) antisymmetric structure constants and the gluon field tensor is
$F^{a\mu\nu}=\partial^{\mu}G^{a\nu}-\partial^{\nu}G^{a\mu}+gf^{abc}G^{b\mu}G^{c\nu}$.
The electromagnetic  Lagrangian term is
$F^{\mu\nu}=\partial^{\mu}A^{\nu}-\partial^{\nu}A^{\mu}$, with $A^{\mu}$ given by (\ref{magfield}). As mentioned above, we  decompose
the gluon field  as in \cite{we11,we12,shakin,shakinn}:
$$
G^{a\mu}={A}^{a\mu}+{\alpha}^{a\mu}
$$
where ${A}^{a\mu}$ and ${\alpha}^{a\mu}$  are the soft and hard gluon  components respectively.
Repeating  the same algebraic steps described in \cite{we11} we rewrite (\ref{lqcduagain}) as the following effective Lagrangian:
$$
\mathcal{L}_{0}={\frac{{m_{G}}^{2}}{2}}{\alpha}^{a}_{0}{\alpha}^{a}_{0}-{\mathcal{B}}_{QCD}
-{\frac{B^{2}}{8\pi}} +\bar{\psi}_{i}^{e}\Big[i\gamma^{\mu}(\delta_{ij}\partial_{\mu}+i\delta_{ij}Q_{e}A_{\mu}
)-\delta_{ij}m_{e}\Big]\psi_{j}^{e}
$$
\begin{equation}
+\sum_{f=u}^{d,s}\bar{\psi}_{i}^{f}\Big\lbrace i\gamma^{\mu}
\Big[\delta_{ij}\partial_{\mu}+i\delta_{ij}Q_{f}A_{\mu}\Big]
+g_{h}\gamma^{0}T^{a}_{ij}\alpha^{a}_{0}-
\delta_{ij}m_{f}\Big\rbrace\psi_{j}^{f}
\label{efqcdl}
\end{equation}
The classical field $\alpha^{a}_{0}$ is the time component of ${\alpha}^{a\mu}$ and it comes from the mean field approximation
${\alpha}^{a}_{\mu}={\alpha}^{a}_{0}\delta_{\mu 0}$ \cite{we11}.
The constant ${\mathcal{B}}_{QCD}$ is the ``bag term'' given by ${\mathcal{B}}_{QCD} \equiv 9{\, \phi_{0}}^{4}/136$ and
${m_{G}}$ is the dynamical gluon mass given by ${m_{G}}^{2}\equiv 9{\, \mu_{0}}^{2}/32$.  The constant
$\mu_{0}$ is an energy scale associated with $\langle A^{2} \rangle$, which is the gluon condensate of dimension two \cite{we11}:
\begin{equation}
 \langle A^{2} \rangle\equiv \langle   g_{s}^{2}{A}^{a\mu} {A}^{b\nu}  \rangle=\langle  g_{s}^2 A^2 \rangle
= -{\frac{\delta^{ab}g^{\mu \nu}}{32}}{\mu_{0}}^{2}
\label{gcdois}
\end{equation}
Since $\langle  g_{s}^2 A^2 \rangle$   $< 0$  we always have ${m_{G}}^{2}$ $>0$.  The constant $\phi_{0}$ is associated with
$\langle F^{2} \rangle$, which is the gluon condensate of dimension four \cite{we11}:
\begin{equation}
{\mathcal{B}}_{QCD} = b \phi_0^4 = \langle  \frac{1}{4} F^{a\mu\nu} F^{a}_{\mu\nu}  \rangle={\frac{\pi^{2}}{g_{s}^{2}}}\langle   {F}^{2}   \rangle
\label{gcqua}
\end{equation}
In the expressions (\ref{efqcdl}) and (\ref{gcdois}) we have two QCD coupling constants given by $g_{h}$ and $g_{s}$.  The coupling $g_{h}$ is associated to the hard gluons, while $g_{s}$ is associated to the soft gluons as in \cite{we11}.

\subsection{Equations of motion and Landau levels}

The following equations of motion are derived from (\ref{efqcdl}):
\begin{equation}
\Big[ i\gamma^{\mu}
\Big(\partial_{\mu}+iQ_{f}A_{\mu}\Big)
+g_{h}\gamma^{0}T^{a}\alpha^{a}_{0}-m_{f}\Big]\psi^{f}=0
\label{psiem}
\end{equation}
\begin{equation}
\Big[ i\gamma^{\mu}
\Big(\partial_{\mu}+iQ_{e}A_{\mu}\Big)
-m_{e}\Big]\psi^{e}=0
\label{psieme}
\end{equation}
\begin{equation}
{m_{G}}^{2}{\alpha}^{a}_{0}=-g_{h}\sum_{f}\rho_{f}^{a}=-g_{h} \, \rho^{a}
\label{azeaem}
\end{equation}
where $\rho^{a}$ is the temporal component of the color vector current $j^{a\nu}$, given by:
 \begin{equation}
j^{a0}=\rho^{a}=\sum_{f}{\bar{\psi}_{i}}^{f}\gamma^{0}T^{a}_{ij}\psi_{j}^{f}
=\sum_{f}{{\psi}_{i}^{\dagger}}^{f}T^{a}_{ij}\psi_{j}^{f}
\label{roqa}
\end{equation}
From the exact solution \cite{bacharia} of the Dirac equation (\ref{psiem}) with magnetic field and hard gluon terms, we have the following
expression for the eigenvalues:
\begin{equation}
\Bigg(E_{\nu}^{f}+g_{h}\mathcal{A}\Bigg)^{2}=m_{f}^{2}+k_{z}^{2}+(2\nu+1)|Q_{f}|B-Q_{f}Bs
\label{prelandquar}
\end{equation}
where $\nu=0,1,2,3,4,5 \dots$ and $s=+1$ or $s=-1$, for the projection {\it{up}} or {\it{down}} of the spin states, respectively. The momentum component
along the magnetic field direction is given by $k_{z}$. As in Ref.  \cite{we11} the constant $\mathcal{A}$ in (\ref{prelandquar}) is the
``algebra valued'' quantity, $\mathcal{A} = c^{\dagger}_{i}T^{a}_{ij}c_{j}\alpha_{0}^{a}$  (with the implicit summation over $i,j=1,2,3$ and $a=1,\dots, 8$),
where  $c_{i}$ is a  color vector, as explained in the Appendix.  In Eq. (\ref{lqcduagain}) and in what follows, we do not include the interaction terms
between the magnetic field and the fermion magnetic moments \cite{magspacks}.  This is because in Ref. \cite{dpm3,magmomentdaryel} it was shown that for  strange
quark matter  in  $\beta-$  equilibrium in  magnetic fields weaker than $10^{18} \, G$ the contribution of these terms can be neglected. More precisely, we will
restrict our  analysis to  $B \leq 5 \times 10^{17} \, G$.

Rescaling the single particle energy as $\tilde{E}_{\nu}^{f} \equiv E_{\nu}^{f}+g_{h}\mathcal{A}$\,,
we are able to rewrite (\ref{prelandquar}) as:
\begin{equation}
\Big(\tilde{E}_{\nu}^{f}\Big)^{2}=m_{f}^{2}+k_{z}^{2}+\Big[2\nu+1-s \times sgn(Q_{f})\Big]|Q_{f}|B
\label{prelandquarrn}
\end{equation}
where $Q_{f}=sgn(Q_{f}) \times |Q_{f}|$.  Defining $2\nu+1-s \times sgn(Q_{f})\equiv 2n$, (\ref{prelandquarrn}) becomes:
\begin{equation}
\tilde{E}_{n}^{f(\pm)}=\pm\sqrt{m_{f}^{2}+k_{z}^{2}+2n|Q_{f}|B}
\label{landquarrn}
\end{equation}
and $n$ denotes the $n^{th}$ Landau level. We note that, except for the rescaling in  $\tilde{E}_{\nu}^{f}$, the equation above is the one
usually found in the literature. Analogously, from the exact solution of (\ref{psieme}) \cite{bacharia}  for the electron  we have:
\begin{equation}
\Big(E_{\nu}^{e}\Big)^{2}=m_{e}^{2}+k_{z}^{2}+(2\nu+1)|Q_{e}|B-Q_{e}Bs
\label{prelande}
\end{equation}
and considering $2\nu+1-s \times sgn(Q_{e})=2n$ we find the energy for the $n^{th}$ Landau level:
\begin{equation}
E_{n}^{e(\pm)}=\pm\sqrt{m_{e}^{2}+k_{z}^{2}+2n|Q_{e}|B}
\label{landquarrne}
\end{equation}

\subsection{Energy density and pressure }

To obtain our EOS we follow the thermodynamical calculations as performed in \cite{dpm3,furn,wenpa}.  The details are in the Appendix, where we also
show the baryon density calculation.
The quark density at zero temperature is given by:
\begin{equation}
\rho=\sum_{f=u}^{d,s}\,{\frac{|Q_{f}|B}{2\pi^{2}}} \, \sum_{n=0}^{n^{f}_{max}}3(2-\delta_{n0}) \, k^{f}_{z,F}(n)
\label{rhoquarksfromomegaendtzero}
\end{equation}
where $k^{f}_{z,F} (n)$ is the quark  Fermi momentum given by:
\begin{equation}
k^{f}_{z,F}(n)=\sqrt{{\nu_{f}}^{2}-m_{f}^{2}-2n|Q_{f}|B}
\label{fermilevelquark}
\end{equation}
The summation over the Landau levels is calculated on the condition that the expression under the square root   in (\ref{fermilevelquark}) is positive,
i.e.,  ${\nu_{f}}^{2} \geq m_{f}^{2}+2n|Q_{f}|B$  \, \cite{dpm3}.  Thus
\begin{equation}
n\leq n^{f}_{max}= {\it{int}} \Bigg[ \, {\frac{{\nu_{f}}^{2}-m_{f}^{2}}{2|Q_{f}|B}} \, \Bigg]
\label{nfmaxquark}
\end{equation}
where $ {\it{int}}[a]$ denotes the integer part of $a$. Analogously, for the electrons  we have:
\begin{equation}
\rho_{e}={\frac{|Q_{e}|B}{2\pi^{2}}} \, \sum_{n=0}^{n^{e}_{max}} (2-\delta_{n0}) \, k^{e}_{z,F}(n)
\label{rhoelessfromomegaendtzero}
\end{equation}
with
\begin{equation}
k^{e}_{z,F}=\sqrt{{\mu_{e}}^{2}-m_{e}^{2}-2n|Q_{e}|B}
\label{fermileveleletron}
\end{equation}
and
\begin{equation}
n\leq n^{e}_{max}={\it{int}} \Bigg[ \, {\frac{{\mu_{e}}^{2}-m_{e}^{2}}{2|Q_{e}|B}} \, \Bigg]
\label{nemaxeletron}
\end{equation}
The energy density, the parallel pressure and the perpendicular pressure are:
$$
\varepsilon={\frac{27}{16}}\,{\xi}^{2}{\rho_{B}}^{2}+{\mathcal{B}}_{QCD}
+{\frac{B^{2}}{8\pi}}
+{\frac{|Q_{e}|B}{2\pi^{2}}} \, \sum_{n=0}^{n^{e}_{max}} (2-\delta_{n0}) \int_{0}^{k^{e}_{z,F}}  dk_{z}\sqrt{m_{e}^{2}+k_{z}^{2}+2n|Q_{e}|B}
$$
\begin{equation}
+\sum_{f=u}^{d,s}{\frac{|Q_{f}|B}{2\pi^{2}}} \, \sum_{n=0}^{n^{f}_{max}} 3(2-\delta_{n0}) \int_{0}^{k^{f}_{z,F}} dk_{z}\sqrt{m_{f}^{2}+k_{z}^{2}+2n|Q_{f}|B}
\label{epsilontempzeromagon}
\end{equation}
\\
$$
p_{\parallel}={\frac{27}{16}}\,{\xi}^{2}{\rho_{B}}^{2}-{\mathcal{B}}_{QCD} -{\frac{B^{2}}{8\pi}}
+{\frac{|Q_{e}|B}{2\pi^{2}}} \, \sum_{n=0}^{n^{e}_{max}} (2-\delta_{n0}) \int_{0}^{k^{e}_{z,F}}  dk_{z}\,{\frac{{k_{z}}^{2}}{\sqrt{m_{e}^{2}+k_{z}^{2}+2n|Q_{e}|B}}}
$$
\begin{equation}
+\sum_{f=u}^{d,s}{\frac{|Q_{f}|B}{2\pi^{2}}} \, \sum_{n=0}^{n^{f}_{max}} 3(2-\delta_{n0}) \int_{0}^{k^{f}_{z,F}} dk_{z} \,{\frac{{k_{z}}^{2}}
{\sqrt{m_{f}^{2}+k_{z}^{2}+2n|Q_{f}|B}}}
\label{parallelpressureaalmendfinaltzero}
\end{equation}
\\
$$
p_{\perp}={\frac{27}{16}}\,{\xi}^{2}{\rho_{B}}^{2}-{\mathcal{B}}_{QCD} + {\frac{B^{2}}{8\pi}}
+{\frac{|Q_{e}|^{2}B^{2}}{2\pi^{2}}} \, \sum_{n=0}^{n^{e}_{max}} (2-\delta_{n0}) n \int_{0}^{k^{e}_{z,F}}  {\frac{dk_{z}}{\sqrt{m_{e}^{2}+k_{z}^{2}+2n|Q_{e}|B}}}
$$
\begin{equation}
+\sum_{f=u}^{d,s}{\frac{|Q_{f}|^{2}B^{2}}{2\pi^{2}}} \, \sum_{n=0}^{n^{f}_{max}} 3(2-\delta_{n0}) n \int_{0}^{k^{f}_{z,F}}   {\frac{dk_{z}}
{\sqrt{m_{f}^{2}+k_{z}^{2}+2n|Q_{f}|B}}}
\label{perppressuerendtzero}
\end{equation}
where  $\xi \equiv g_{h}/m_{G}$, as in \cite{we12}. Throughout this work we compute the values for the baryon density $\rho_{B}$ as multiples of the
usual nuclear matter $\rho_{0}=0.17\, fm$.

\begin{figure}
\begin{center}
\subfigure[ ]{\label{fig1a}
\includegraphics[width=0.48\textwidth]{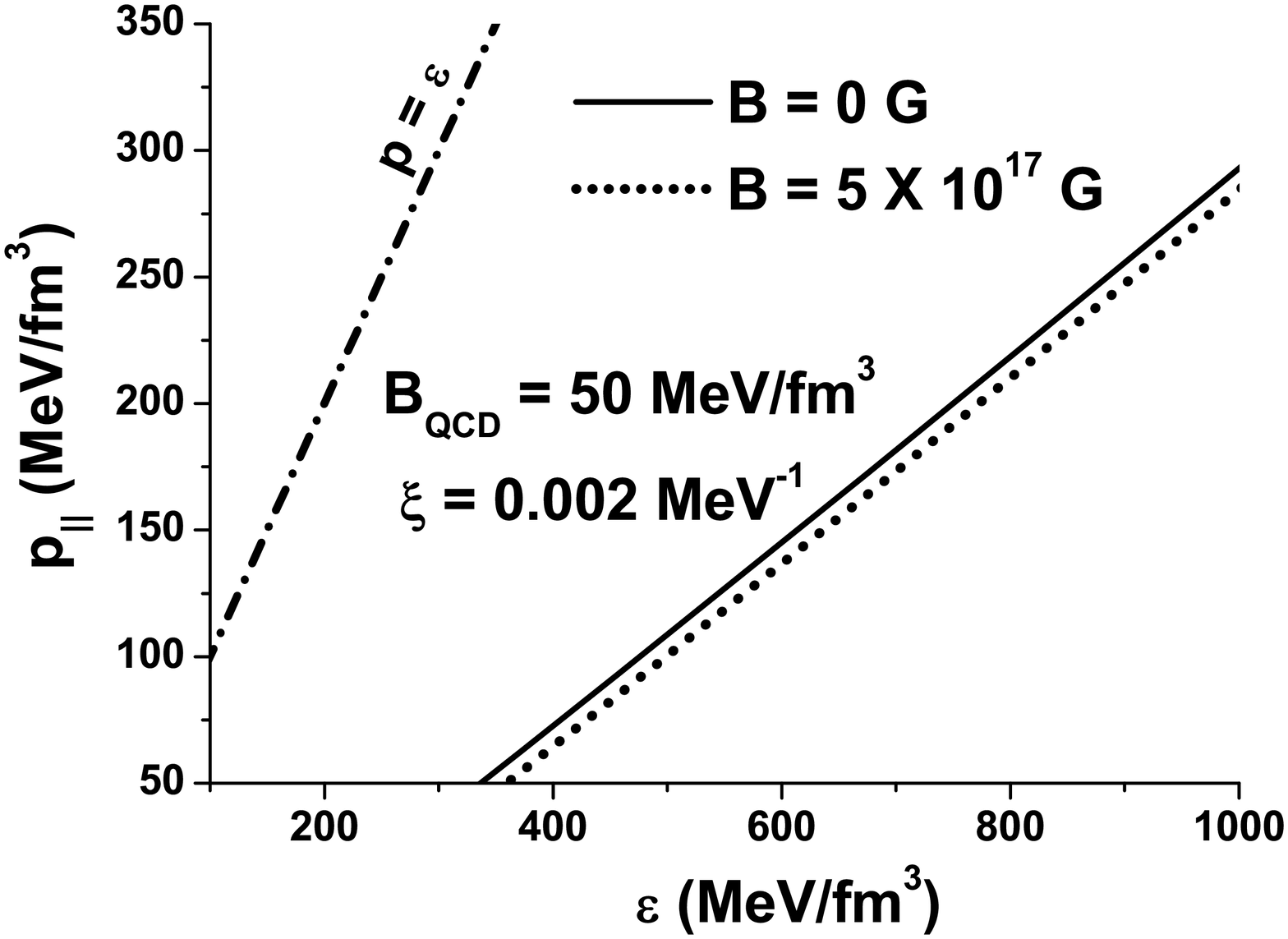}}
\subfigure[ ]{\label{fig1b}
\includegraphics[width=0.48\textwidth]{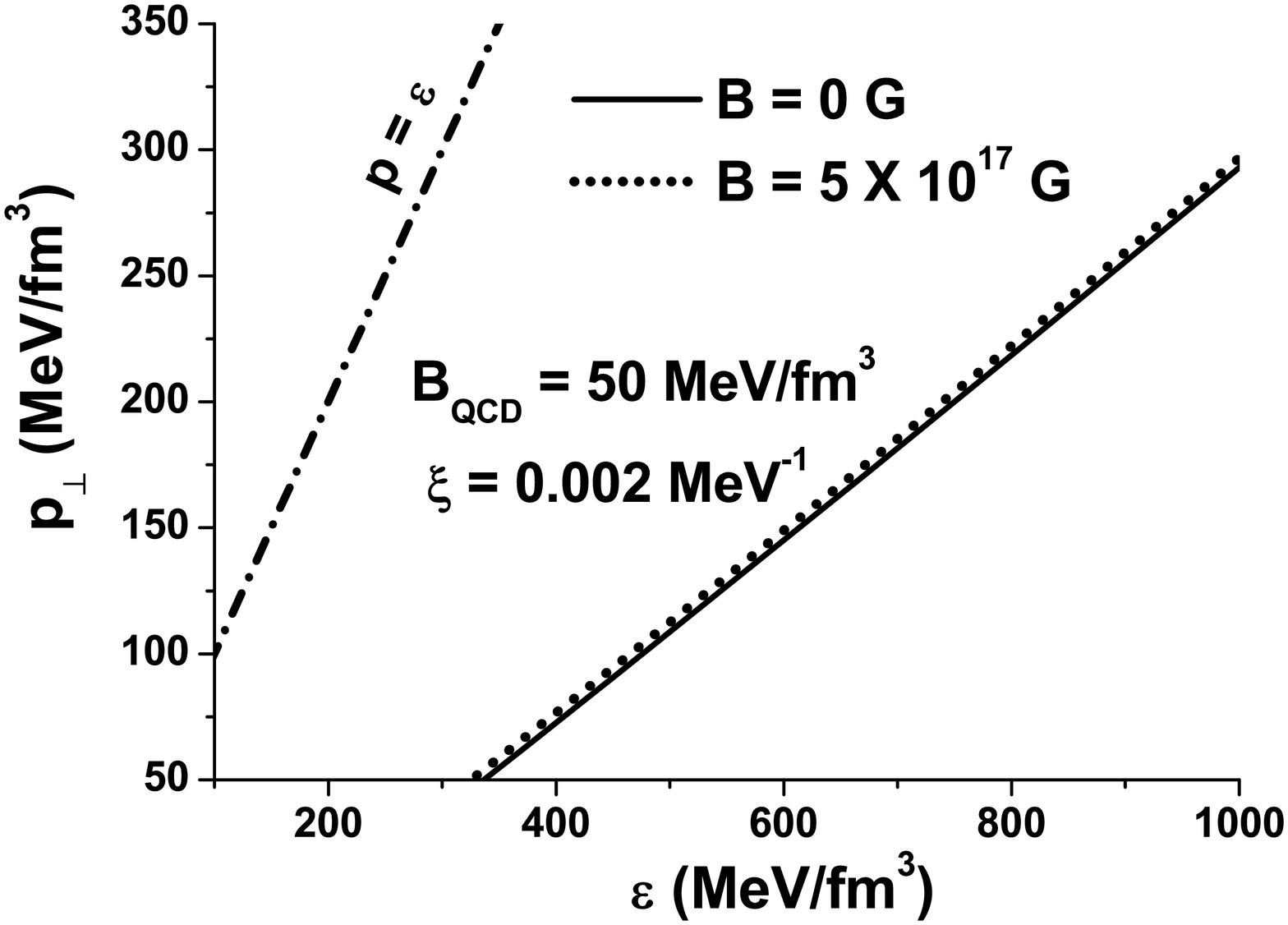}}
\end{center}
\caption{Dependence of the EOS on the magnetic field. (a) Parallel pressure. (b) Perpendicular pressure.}
\label{eoss}
\end{figure}

\begin{figure}[h]
\epsfig{file=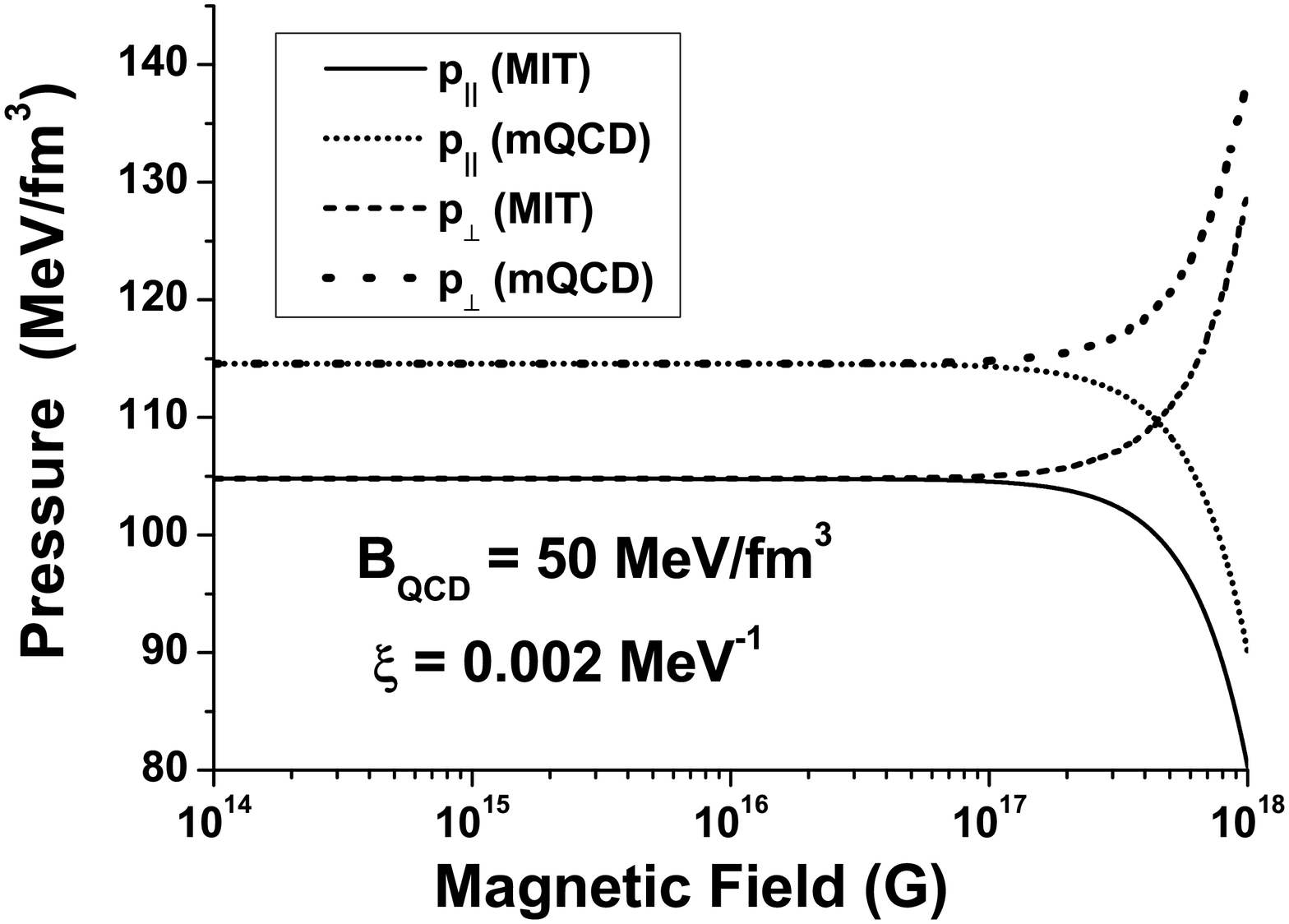,width=135mm}
\caption{The splitting between the paralell and perpendicular pressures as a function of the magnetic field for mQCD and for the MIT bag model.}
\label{split}
\end{figure}

We remember that when $\xi=0$ we recover the result of the MIT bag model.
In this case we do not consider the electrons and just focus on the pure QCD matter, varying the baryon density  from $1.3 \,  \rho_{0}$ to $4.8 \, \rho_{0}$.
The results are shown in Fig. \ref{eoss} for the  parallel and perpendicular pressures.  We have chosen
$\xi=0.002 \, MeV^{-1}$, ${\mathcal{B}}_{QCD} = 50 \, MeV/fm^{3}$ and varied the magnetic field from zero to  $B=5 \times 10^{17} \, G$.
As can be seen in the figure there are no causality violations (in which case we would have ${c_{s}}^{2} = \partial p / \partial \varepsilon > 1$).
The parallel pressure (Fig. \ref{eoss}a) decreases as  the magnetic field increases, while the perpendicular pressure (Fig. \ref{eoss}b) increases with
the magnetic field. Up to the considered maximum value of the magnetic field, the dependence of $p_{\parallel}$ and $p_{\perp}$  with $B$ is very mild.
Moreover they are almost equal to each other. However at higher values of the magnetic field there is a rapid splitting between $p_{\parallel}$ and
$p_{\perp}$,  which is  shown in  Fig. \ref{split}. At $B = 5 \times 10^{17} \, G$  the difference between  the two pressures is not yet very pronounced
(less than 10 \%) and the spherical symmetry  can still be used to derive the standard Tolman-Oppenheimer-Volkov equations.  The results shown in  Fig. \ref{split} are compatible with those
shown in Fig. 1 of Ref. \cite{laura} and also with those shown in Fig. 5 of Ref. \cite{dex-12}, where the quark matter was represented by slightly different versions
of  the MIT bag model equation of state.  The main difference is that,  while in these works the pressure anisotropy starts at $ B \simeq 10^{18}$ G, in our
calculations it starts earlier, at $B \simeq 10^{17}$ G. This happens because of the term proportional to ${\rho_B}^2$ appearing in
the energy density (\ref{epsilontempzeromagon}) and in the pressure (\ref{parallelpressureaalmendfinaltzero})-(\ref{perppressuerendtzero}),
which depends quadratically on $B$, as can be seen from (\ref{rhoquarksfromomegaendtzero}). This term anticipates the high $B$ behavior of the pressure and
the appearance of the pressure anisotropy.

At this point a remark is in order. As discussed in detail in  \cite{dex-12}, there is a controversy in the literature concerning the existence or non-existence
of pressure anisotropies. In early works (see the references quoted in \cite{dex-12}) it was explicitly demonstrated that, in the presence of a background
magnetic field, a Fermi-gas of spin-one-half particles possesses a pressure anisotropy. Later the calculations were revisited and the effects of the anomalous
magnetic moment were included.  It was concluded that the pressure anisotropy exists for both charged and uncharged particles, with and without anomalous
magnetic moment. On the other hand, in Ref. \cite{bland} and more  recently in \cite{yako} it was  argued that, due to the presence of a non-vanishing magnetization
one needed to additionally take into account the Lorentz force of the external magnetic field on the bound current densities, which would lead the system to
isotropization.  While this question is certainly very interesting we will stay on the safe side, avoid the region of very high $B$ and consider only values of the
magnetic field where the pressure is isotropic. More precisely, in what follows we will compute the star masses up to  $B = 5 \times 10^{17} \, G$  using  the two
different pressures and interpret the results as upper and lower limits of our calculations, regarding their difference as a theoretical error.

\section{Stability conditions}

We wish to study  stellar models with stable strange quark matter (described by the mQCD equation of state) and hence we will impose the stability conditions.
The first  condition  is the  existence of chemical equilibrium in the weak processes involving the quarks $u$, $d$,
$s$ and electrons \cite{farhi,glend}:
$$
u+e^{-} \rightarrow d + \nu_{e},\hspace{1.5cm}
u+e^{-} \rightarrow s + \nu_{e},
$$
\begin{equation}
d \rightarrow u+e^{-} + \bar{\nu}_{e},\hspace{1.5cm}
s \rightarrow u+e^{-} + \bar{\nu}_{e},  \hspace{1cm} \textrm{and} \hspace{1cm}
s + u\rightarrow d+ u.
\label{w}
\end{equation}
which provides the following relations among the chemical potentials:
\begin{equation}
\nu_{d} = \nu_{s} \equiv \mu
\hspace{1cm} \textrm{and} \hspace{1cm}
\nu_{u} + \mu_{e}= \mu
\label{ce}
\end{equation}
The second condition is the global charge neutrality enforced by:
\begin{equation}
{\frac{2}{3}}\rho_{u}={\frac{1}{3}}\rho_{d}+{\frac{1}{3}}\rho_{s}+\rho_{e},
\label{cn}
\end{equation}
The third condition is the baryon number conservation, which implies that \cite{we12}:
\begin{equation}
\rho_{B}={\frac{1}{3}}\rho={\frac{1}{3}}(\rho_{u}+\rho_{d}+\rho_{s})
\label{bc}
\end{equation}
The last condition is the requirement that the energy per baryon must be lower than the infinite baryonic matter defined in \cite{farhi} and higher
than the two flavor quark matter at the ground state \cite{farhi}. We must impose that \cite{we12}:
\begin{equation}
{\frac{\varepsilon}{\rho_{B}}}\bigg{|}_{(3\,\textrm{-flavor})}   \,\,\,\,\, \leq  \,\, 934 \,\, \mbox{MeV}  \,\,   \leq
\,\,\,\,\,  \frac{\varepsilon}{\rho_{B}}\bigg{|}_{(2\,\textrm{-flavor})}
\label{stable}
\end{equation}

\begin{figure}[h]
\epsfig{file=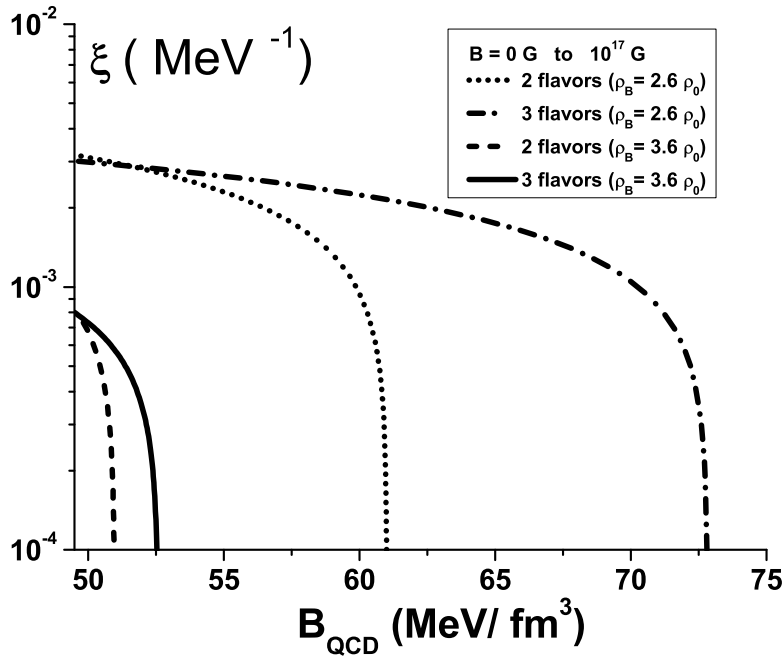,width=135mm}
\caption{Stability windows defined by the conditions (\ref{ce}) to (\ref{stable}).}
\label{figstablewindows1}
\end{figure}

We find numerically the values of $\xi$ and $\mathcal{B}_{QCD}$ which satisfy (\ref{ce}) to (\ref{stable}) simultaneously.  Some examples of  stability regions
in the  $\xi \, - \,  \mathcal{B}_{QCD}$ parameter space are presented in Fig. \ref{figstablewindows1}, where the regions defined  by  the curves are the
``stability windows'' of $\xi$ as a function of $\mathcal{B}_{QCD}$. We observe that increasing the baryon density the window ``shrinks'', i.e.,  the
stability area  becomes smaller and  thinner. There is a maximum baryon density, $\rho_B \simeq 3.7 \, \rho_0$,
beyond which there is no stability window. This was expected and could be anticipated by looking at the first term of Eq. (\ref{epsilontempzeromagon}). When
$\rho_B$ grows this term becomes dominant and the ratio   $\varepsilon / \rho_B$ grows in such a way that it can no longer satisfy the left inequality in
(\ref{stable}).  The same reasoning applies to the value of the magnetic field. From (\ref{epsilontempzeromagon}) we can infer that there is a value of $B$, beyond
which there will be no stability.

\begin{figure}[h]
\epsfig{file=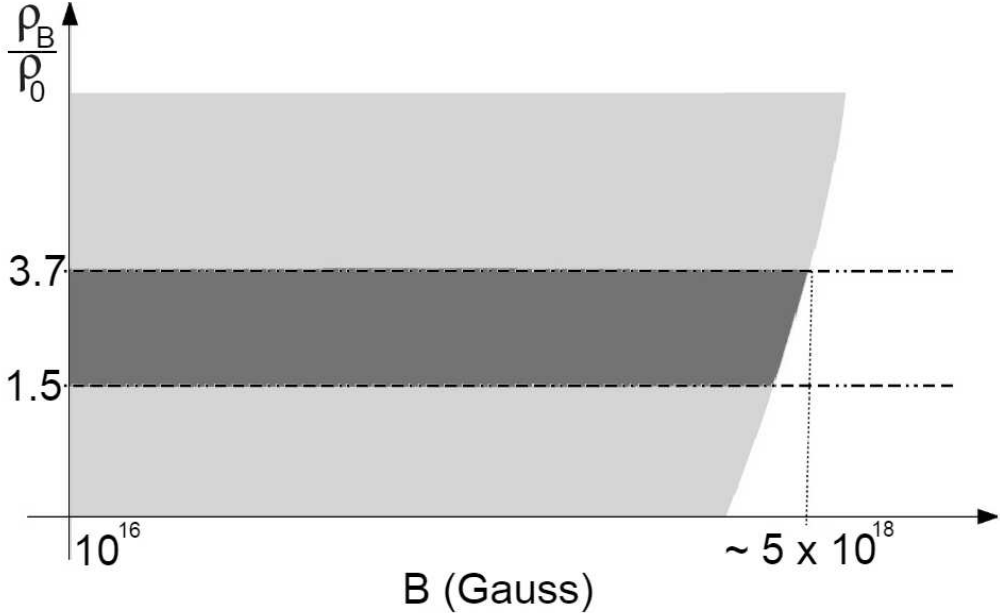,width=145mm}
\caption{Stability diagram: baryon density ratio as function of magnetic field. The points in the light grey area satisfy the
conditions (\ref{ce}), (\ref{cn}) and (\ref{bc}).  Points in the dark grey area satisfy also the condition (\ref{stable}).}
\label{stabquali}
\end{figure}

In Fig. \ref{stabquali} we show the diagram of stability as function of the magnetic field for the mQCD equation of state. From the figure we can observe that
there is a maximum values of $\rho_B$ and of the field $B$, beyond which there is no stability.

\section{Stellar structure}

As usual, to describe the structure of a static and non-rotating compact star, the Einstein equations are solved for the spherical, isotropic,
static and  general relativistic ideal fluid in hydrostatic equilibrium. Under these conditions, the solution of the Einstein equations provides  the
Tolman-Oppenheimer-Volkoff (TOV) equation for the pressure $p(r)$:
\begin{equation}
\frac{dp}{dr}=-\frac{G \epsilon(r) M (r)}{r^2} \left[ 1 + \frac{p(r)}{\epsilon(r)} \right] \left[ 1 + \frac{4\pi r^3 p(r)}{M(r)} \right] \times
\left[ 1 - \frac{2GM(r)}{r} \right]^{-1},
\label{tovagain}
\end{equation}
where $G$ is the Newton gravitational constant. The mass $M(r)$ of the compact star is given by the mass continuity equation:
\begin{equation}
\frac{dM(r)}{dr}=4\pi r^2\epsilon(r).
\label{massagain}
\end{equation}

In general, magnetic fields tend to deform a star and for larger  magnetic fields, there will be a  large
deformation caused by changes in the metric inside the star. This aspect has been well studied  in the
literature, as for example, in \cite{tovsno}. In these situations, where the magnetic fields are of the order of  $10^{18}\, G$,
the use of the spherically symmetric TOV equations to study  the star structure is not appropriate.  Therefore we will restrict our study to fields up to $5 \times 10^{17}\, G$.

We solve numerically the coupled nonlinear equations (\ref{tovagain}) and (\ref{massagain}) for $p(r)$ and $M(r)$, in order to obtain the mass-radius
diagram.  The possible magnetic field effects in the stellar structure come from the EOS.
We consider the central energy density $\epsilon(r = 0)=\epsilon_{c}$ and then we integrate both (\ref{tovagain}) and (\ref{massagain}) from $r=0$
up to $r=R$ (stellar radius), where the pressure at the surface is zero: $p(r=R)=0$.

\begin{figure}[h]
\epsfig{file=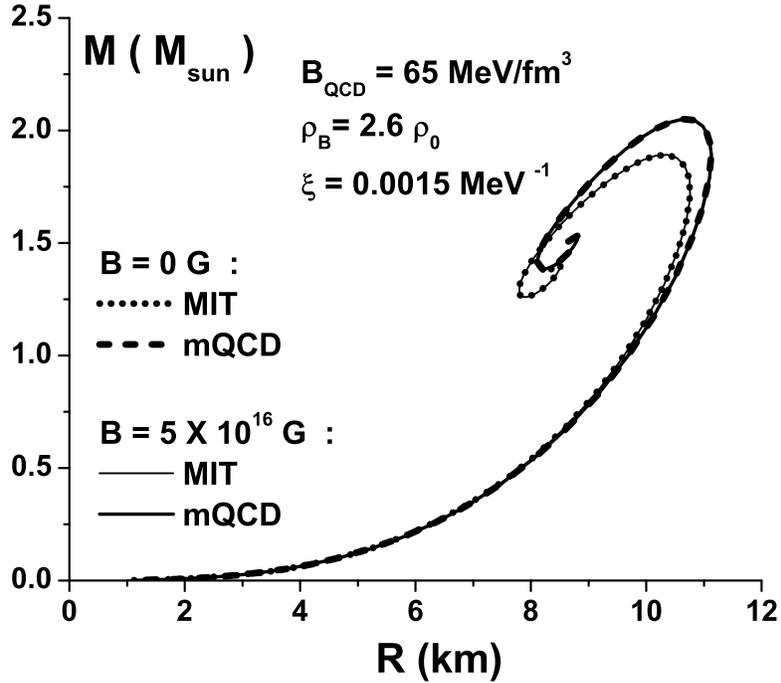,width=135mm}
\caption{Mass-radius diagrams.  Two values of the magnetic fields with $\mathcal{B}_{QCD}$ and $\xi$ allowed by the stability
conditions at $\rho_{B}=2.6 \, \rho_{0}$. The largest masses are $2.05$ (mQCD) and $1.89$  (MIT).
In these cases $p_{\parallel}=p_{\perp}$ which permits the use of TOV.}
\label{tovcomp}
\end{figure}

In Fig. \ref{tovcomp}  we present some solutions of the TOV equations and the resulting  mass-radius diagrams. We fix $\mathcal{B}_{QCD}$, $\xi$ and $\rho_B$
respecting the stability windows and consider two values for the magnetic field. From the figure we observe that the mQCD model predicts larger masses than the MIT one and also that all values of the magnetic field, from zero to $5 \times 10^{16}\,G$,  yield the same mass-radius curves.
In this range of $B$ values the parallel and perpendicular pressures are equal. One of main conclusions of Ref. \cite{we12} was that, with the EOS provided by mQCD, it was possible to have strange quark stars with two (or more) solar masses. The purpose of the calculations presented in Fig. \ref{tovcomp}  is to show that this remains true for strong magnetic fields.

\begin{figure}
\begin{center}
\subfigure[ ]{\label{fig6a}
\includegraphics[width=0.547\textwidth]{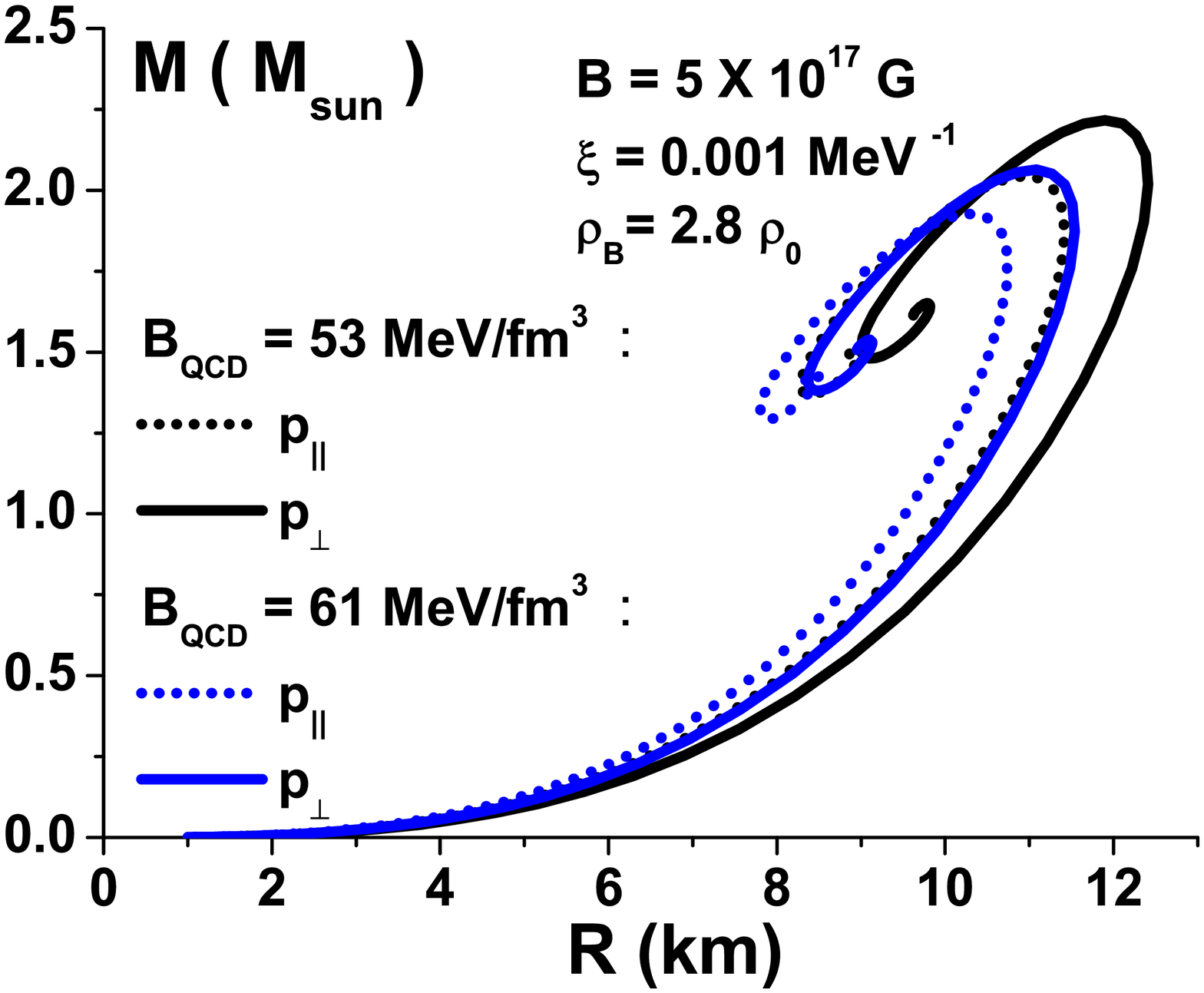}}
\hskip-2.1cm
\subfigure[ ]{\label{fig6b}
\includegraphics[width=0.547\textwidth]{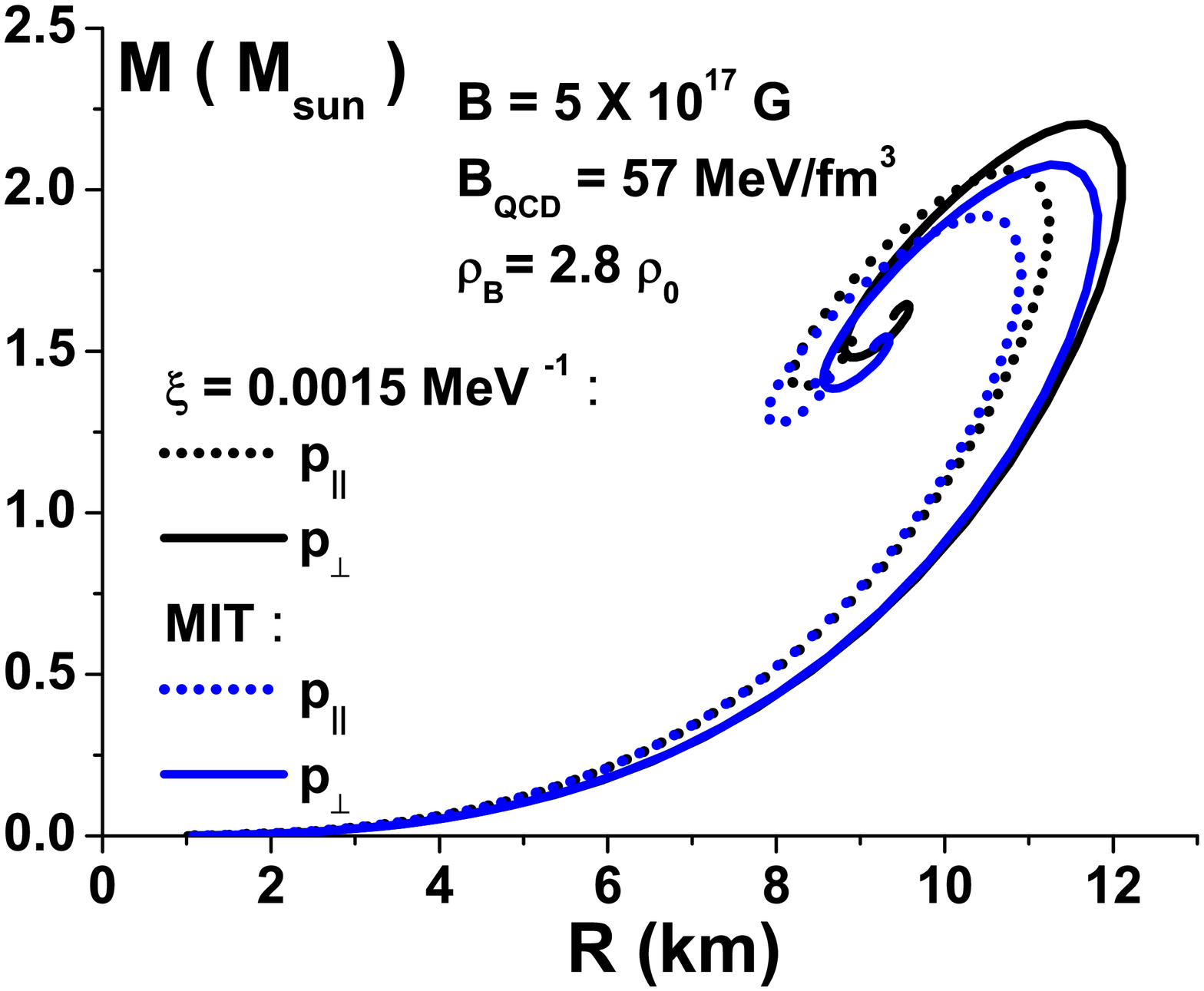}}
\end{center}
\vskip-0.5cm
\caption{Mass-radius diagram for a fixed value of the magnetic field and the baryon density with $p_{\parallel}<p_{\perp}$.
(a) Fixed $\xi$. For $\mathcal{B}_{QCD}=53 \, MeV/fm^{3}$ the largest masses are $2.22$ and $2.04$ (along the dotted lines), calculated with the
perpendicular and parallel pressures respectively.  Analogously for $\mathcal{B}_{QCD}=61 \, MeV/fm^{3}$ the largest masses are $2.06$ and $1.93$
(along the solid lines). (b) Fixed  $\mathcal{B}_{QCD}$. For $\xi=0.0015 \, MeV^{-1}$ the largest masses are $2.20$ and and $2.06$ (along the dotted
lines), calculated with  perpendicular and parallel pressures respectively. Analogously for MIT $(\xi=0 \, MeV^{-1})$ the largest masses are $2.08$ and $1.92$ (along the solid lines).}
\label{tov5X10a17}
\end{figure}

\begin{figure}[h]
\epsfig{file=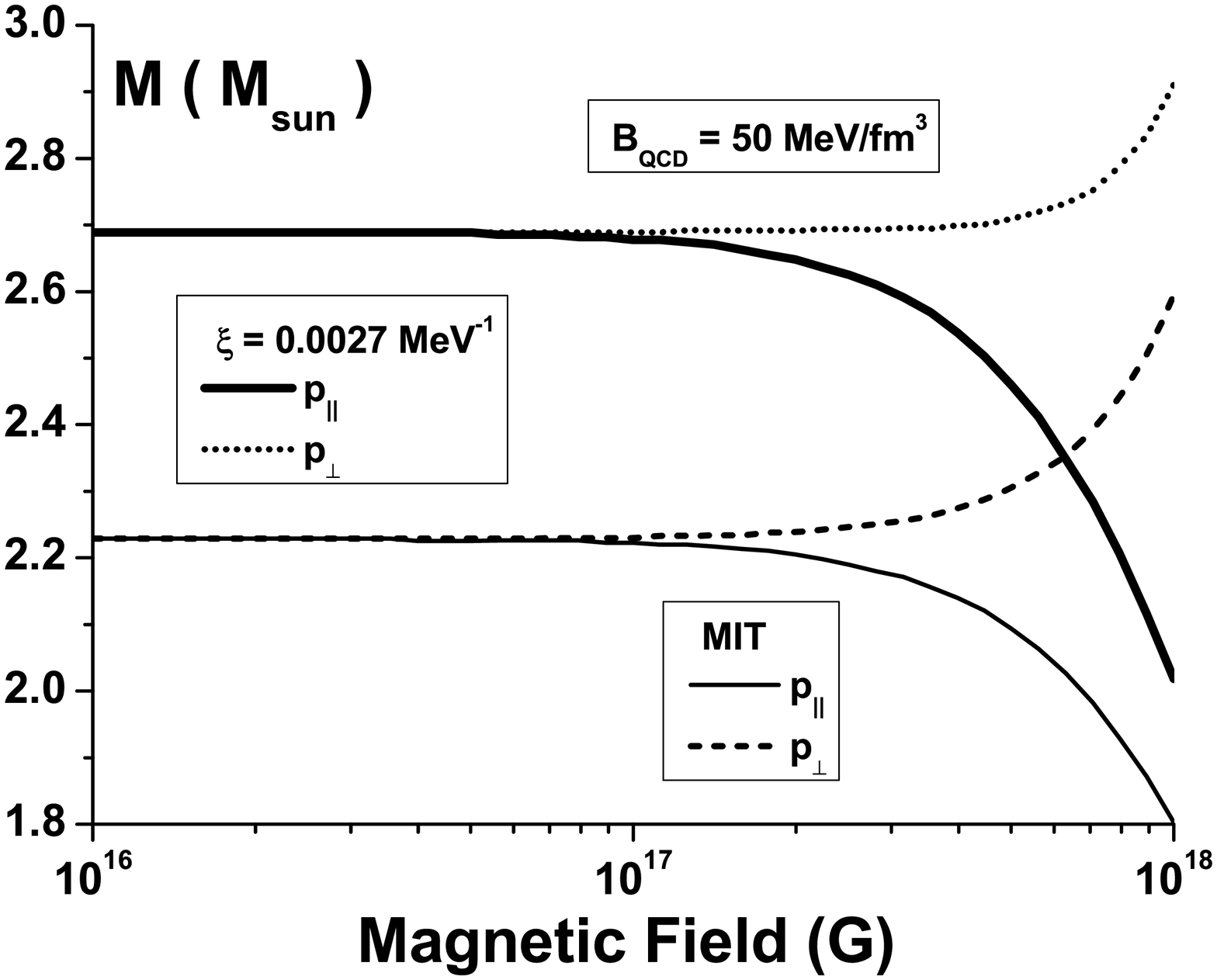,width=135mm}
\caption{Effect of the pressure anisotropy on the star masses computed with the TOV equations. The chemical potentials obey the stability conditions
(\ref{ce}), (\ref{cn}), (\ref{bc}) and (\ref{stable}), given by $\nu_{u}=300\,MeV$, $\nu_{d}=\nu_{s}=316.5\,MeV$ and $\mu_{e}=16.5\,MeV$ . }
\label{behavior}
\end{figure}

In Fig. \ref{tov5X10a17} we show how the mass-radius curves change when we  keep the magnetic field constant and change $\mathcal{B}_{QCD}$ and  $\xi$.
We solve the equations (\ref{tovagain}) and (\ref{massagain})  using the parallel and perpendicular pressures, given respectively
by (\ref{parallelpressureaalmendfinaltzero}) and (\ref{perppressuerendtzero}). As expected, smaller values of $\mathcal{B}_{QCD}$ imply higher
pressure and higher masses, as we can see in Fig.  \ref{tov5X10a17}a.  A larger value of $\xi$ increases the pressure and the values
of the obtained masses, as shown in Fig.  \ref{tov5X10a17}b.  These results are in qualitative agreement with those found in \cite{we12} at zero magnetic field.
At $B = 5 \times 10^{17}$ G there is a visible difference between the results obtained with parallel and perpendicular pressures. This is the point where  we stop our calculations and the difference between the results obtained with $p_{\parallel}$ and  $p_{\perp}$ give an estimate of our theoretical error.
However, if we insist on solving the TOV equations even for values of $B$  for which the pressure is anisotropic,  we obtain the masses shown in
Fig. \ref{behavior}.  Comparing with Fig. \ref{split}, we observe that  there is a direct correspondence between pressure and star mass.
Under the same change of $B$ (from $10^{17}$ to $10^{18}$ Gauss)  the pressures and masses change by a similar  amount of $\simeq 20$ \% or less.

\section{Conclusions}

We have evaluated the  equation of state derived in  Ref. \cite{we11} (which we call mQCD) at very large baryon densities, where deconfined quark matter
should exist. In our model the ideal gas behavior is reached in the limit $\mathcal{B}_{QCD} \rightarrow 0$, $g_h \rightarrow 0$ and $m_{G} \rightarrow 0$
(respecting the condition $\xi = g_h/m_{G} \rightarrow 0$). In this limit we obtain the Stefan-Boltzmann (SB) equation of state. The results of
Ref. \cite{fkv16} suggest that the SB limit is not yet reached at chemical potentials in the range $1 \,  \mbox{GeV} \leq \mu_B \leq 3 \, \mbox{GeV}$ .
In our model this means that the pressure is lowered (with respect to the SB value) because the quarks have non-zero masses  or because of a non-vanishing
gluon condensate or because of the two reasons combined.

We have introduced the magnetic field in the mQCD equation of state. We observe the splitting of the pressure into parallel, $p_{\parallel}$, and perpendicular,
$p_{\perp}$,  pressures. When $B$ increases, $p_{\perp}$ increases whereas $p_{\parallel}$ decreases.
In our model this splitting starts to happen when $B \simeq 10^{17}$ G  and it is a modest effect until $B  \simeq 5 \times 10^{17}$ G.
Since  larger pressures are essential to generate stars with larger masses, it is not clear a priori what is the effect of the magnetic field on the mass of the star. Moreover, increasing $B$ the stability window shrinks and from $B \simeq 5 \times 10^{18}$ G on, we can not find any stable quark star.
At these values the difference between  $p_{\parallel}$ and $p_{\perp}$ is so large that we should no longer use the standard spherically symmetric TOV equations.
Even though it was not possible to determine a clear trend of mass-radius curves with the magnetic field, we could find stars with more than two solar masses at $B \simeq 10^{17} \, G$. From this we can conclude that the heavy  and magnetized stellar objects mentioned in the introduction can be, among other
possibilities, quark stars.

To summarize: in our model the magnetic field does not generate any noticeable effect until $10^{17}$ G. From this point on,  it generates a pressure
anisotropy which precludes the use of the TOV equations. Moreover, it rapidly increases the energy density closing the stability window for this kind of strange quark matter.

\newpage

\section{Appendix}

In this Appendix we present some details of the derivation of Eqs. (\ref{rhoquarksfromomegaendtzero}), (\ref{epsilontempzeromagon}),
(\ref{parallelpressureaalmendfinaltzero})  and  (\ref{perppressuerendtzero}).
\subsection{Baryon density}

The $c_{j}$ is the quark color vector used in some textbooks \cite{grif}:
\begin{equation}
c_{1}=
\left( \begin{array}{c}
1 \\
0  \\
0
\end{array} \right)
\hspace{0.3cm} \textrm{for red,} \hspace{0.2cm}
c_{2}=
\left( \begin{array}{c}
0 \\
1  \\
0
\end{array} \right)
\hspace{0.3cm} \textrm{for blue,} \hspace{0.2cm}
c_{3}=
\left( \begin{array}{c}
0 \\
0  \\
1
\end{array} \right)
\hspace{0.3cm} \textrm{for green}
\label{qcolor}
\end{equation}
From the above definitions it follows that $ c^{\dagger }_{i}\delta_{ij}c_{j}=c^{\dagger}_{1}c_{1}+c^{\dagger}_{2}c_{2} +c^{\dagger}_{3}c_{3}=3$.
For future purposes we will replace the above sum by the following average:
\begin{equation}
c^{\dagger }_{i}\delta_{ij}c_{j} \rightarrow {\frac{c^{\dagger }_{i}\delta_{ij}c_{j}}{(\textrm{number of quark colors})}}
={\frac{c^{\dagger}_{1}c_{1}+c^{\dagger}_{2}c_{2}
+c^{\dagger}_{3}c_{3}}{3}}=1
\label{qc1s}
\end{equation}
With the help of  (\ref{qcolor}) we are able to calculate the relation between  $\rho^{a}$ previously identified in (\ref{roqa}) and the net quark density
$\rho$. We perform the product $\rho^{a}\rho^{a}$ taking the average over the number of $SU(3)$ generators, which is $8$, as follows:
$$
\rho^{a}\rho^{a}=\sum_{f}\rho_{f}^{a}\sum_{f'}\rho_{f'}^{a} \longrightarrow \langle \sum_{f}\rho_{f}^{a}\sum_{f'}\rho_{f'}^{a} \rangle
={\frac{1}{8}}\sum_{f}\rho_{f}^{a}\sum_{f'}\rho_{f'}^{a}
$$
$$
={\frac{1}{8}}\sum_{f}({\psi}^{\dagger \, f}_{i}T^{a}_{ij}\psi^{f}_{j})
\sum_{f'}({\psi}^{\dagger \, f'}_{k}T^{a}_{kl}\psi^{f'}_{l})
={\frac{1}{8}}\sum_{f}({c}^{\dagger}_{i}T^{a}_{ij}c_{j}){\psi}^{\dagger \, f}\psi^{f}
\sum_{f'}({c}^{\dagger}_{k}T^{a}_{kl}c_{l}){\psi}^{\dagger \, f'}\psi^{f'}
$$
The result $({c}^{\dagger}_{i}T^{a}_{ij}c_{j})({c}^{\dagger}_{k}T^{a}_{kl}c_{l})=3$ is obtained from the Gell-Mann matrices and from the color vectors
(\ref{qcolor}):
$$
\rho^{a}\rho^{a}=\sum_{f}\rho_{f}^{a}\sum_{f'}\rho_{f'}^{a} =  {\frac{3}{8}}\sum_{f}({\psi}^{\dagger \, f}\psi^{f})
\sum_{f'}({\psi}^{\dagger \, f'}\psi^{f'})
$$
As $\sum_{f}({\psi}^{\dagger \, f}\psi^{f})=\sum_{f}\,\rho_{f}=\rho$, where $f = u,d,s$ and  $\rho$ is the total net quark density, we have:
\begin{equation}
\rho^{a}\rho^{a}={\frac{3}{8}}{\rho}^{2}
\label{roqaa}
\end{equation}
The baryon density $\rho_{B}$ is related to net quark density through:
\begin{equation}
\rho_{B}={\frac{1}{3}}\rho
\label{robao}
\end{equation}

\subsection{Thermodynamical quantities}

Performing the calculations presented in \cite{dpm3,furn,wenpa} and starting from (\ref{efqcdl}) we arrive at the following thermodynamical potential:

$$
\Omega=\Bigg[
-{\frac{{m_{G}}^{2}}{2}}{\alpha}^{a}_{0}{\alpha}^{a}_{0}+{\mathcal{B}}_{QCD} + {\frac{B^{2}}{8\pi}} \Bigg]V
$$
\begin{equation}
+T\sum_{{\vec{k},s,n}} \Bigg\{ ln\Big(1-d_{e}\Big) + ln\Big(1-\bar{d}_{e}\Big)  \Bigg\}
+T\sum_{f=u}^{d,s}\,\sum_{{\vec{k},s,n}} \Bigg\{ ln\Big(1-d_{f}\Big) + ln\Big(1-\bar{d}_{f}\Big)  \Bigg\}
\label{termopotfrwn}
\end{equation}
where $V$ is the volume and $T$ is the temperature.  The fermion distribution functions are:
\begin{equation}
d_{i}\equiv{\frac{1}{1+e^{(\mathcal{E}^{i}_{n}-\nu_{i})/T}}}
\hspace{1.3cm} \textrm{and} \hspace{1.3cm}
\bar{d}_{i}\equiv{\frac{1}{1+e^{(\mathcal{E}^{i}_{n}+\nu_{i})/T}}}
\label{dists}
\end{equation}
with $i=e$ for the electron and $i=f$ for each quark.
The $\nu_{e}$ is the chemical potential for the electrons and the effective chemical potential of the quark $f$ is defined as:
$\nu_{f} \equiv \mu_{f} +{g_{h}}{({c}_{i}^{\dagger}}T^{a}_{ij}c_{j})\alpha^{a}_{0}= \mu_{f} +{g_{h}}\mathcal{A}$.
From (\ref{landquarrne}) the energy of the electron is:
\begin{equation}
\mathcal{E}^{e}_{n}=\sqrt{m_{e}^{2}+k_{z}^{2}+2n|Q_{e}|B}
\label{energtotaleles}
\end{equation}
and using (\ref{landquarrn}) in the evaluation of (\ref{termopotfrwn}) the energy of the quark $f$ is now defined as:
\begin{equation}
\mathcal{E}^{f}_{n}=\sqrt{m_{f}^{2}+k_{z}^{2}+2n|Q_{f}|B}
\label{energtotalnfs}
\end{equation}
For a magnetic field pointing along the $z$ direction, the momentum of a charged particle is restricted to discrete
Landau levels \cite{dpm3,mags,magspacks,magmomentdaryel} and hence:
$$
{\frac{S}{(2\pi)^{2}}}\int_{-\infty}^{\infty}\int_{-\infty}^{\infty} dk_{x} \, dk_{y}={\frac{S|Q_{i}|B}{2\pi}}
$$
with $S$ being the area in the $x-y$ plane.  From this last expression we have:
\begin{equation}
\int_{-\infty}^{\infty}\int_{-\infty}^{\infty} dk_{x} \, dk_{y}=2\pi|Q_{i}|B
\label{precontlim}
\end{equation}
and the statistical sum becomes:
\begin{equation}
{\frac{1}{V}}\sum_{\vec{k},s,n} \longrightarrow{\frac{1}{(2\pi)^{3}}}
 \sum_{n}\,\gamma_{i}(n)\,\int d^{3}k=
{\frac{|Q_{i}|B}{(2\pi)^{2}}} \sum_{n}\,\gamma_{i}(n)\,\int_{-\infty}^{\infty} dk_{z}
\label{contlim}
\end{equation}
where $\,\gamma_{i}(n)\,$ is the statistical degeneracy factor of the $i^{th}$ fermion. For the electron we have
$\gamma_{e}(n)=(2-\delta_{n0})$ and for each quark $f$ we have $\gamma_{f}(n)=3 \, (2-\delta_{n0})$, where the numerical factor ``$3$'' is due the color.
The  pressure parallel to the magnetic field $(p_{\parallel})$, the magnetization $(M)$
and the perpendicular pressure $(p_{\perp})$ are given respectively by \cite{dpm3,mags}:
\begin{equation}
p_{\parallel}=-{\frac{\Omega}{V}}
\hspace{0.8cm} \textrm{,}  \hspace{1cm}
M=-{\frac{1}{V}}{\frac{\partial \Omega}{\partial B}}={\frac{\partial p_{\parallel}}{\partial B}}
\hspace{1cm} \textrm{and}  \hspace{1cm}
p_{\perp}=p_{\parallel}-MB
\label{pressuresM}
\end{equation}
The electron density $\rho_{e}$, the quark density $\rho$ and the entropy density $s$ read \cite{furn,wenpa}:
\begin{equation}
\rho_{e}=-{\frac{1}{V}}{\frac{\partial \Omega}{\partial \mu_{e}}}
\textrm{\, ,}  \hspace{1cm}
\rho=-{\frac{1}{V}}{\frac{\partial \Omega}{\partial \mu_{f}}}
\hspace{1cm} \textrm{and}  \hspace{1cm}
s=-{\frac{1}{V}}{\frac{\partial \Omega}{\partial T}}
\label{rhoqs}
\end{equation}
The energy density $\varepsilon$ is calculated from the Gibbs relation \cite{furn,wenpa}:
\begin{equation}
\varepsilon=-p_{\parallel}+Ts+\sum_{f}\mu_{f}\rho_{f}
\label{energydensitytotal}
\end{equation}
The evaluation of (\ref{pressuresM}) to (\ref{energydensitytotal}) with the potential (\ref{termopotfrwn}) gives the following results:
$$
p_{\parallel}={\frac{3{g_{h}}^{2}}{16{m_{G}}^{2}}}\rho^{2}-{\mathcal{B}}_{QCD}- {\frac{B^{2}}{8\pi}}
+{\frac{|Q_{e}|B}{2\pi^{2}}} \sum_{n}(2-\delta_{n0})\int_{0}^{\infty} dk_{z} \,{\frac{{k_{z}}^{2}}{\mathcal{E}^{e}_{n}}}\, \Big(d_{e}+\bar{d}_{e} \Big)
$$
\begin{equation}
+\sum_{f=u}^{d,s}\, {\frac{|Q_{f}|B}{2\pi^{2}}} \sum_{n}3(2-\delta_{n0})\int_{0}^{\infty} dk_{z} \,{\frac{{k_{z}}^{2}}{\mathcal{E}^{f}_{n}}}\,
\Big(d_{f}+\bar{d}_{f} \Big)
\label{parallelpressurefinal}
\end{equation}
$$
M=-B-T \, {\frac{|Q_{e}|}{2\pi^{2}}} \sum_{n}(2-\delta_{n0})\int_{0}^{\infty} dk_{z}\Bigg[ ln(1-d_{e}) + ln(1-\bar{d}_{e})\Bigg]
$$
$$
-T \sum_{f=u}^{d,s}\, {\frac{|Q_{f}|}{2\pi^{2}}} \sum_{n}3(2-\delta_{n0})\int_{0}^{\infty} dk_{z} \Bigg[ ln(1-d_{f}) + ln(1-\bar{d}_{f})\Bigg]
$$
$$
-{\frac{|Q_{e}|B}{2\pi^{2}}} \sum_{n}(2-\delta_{n0})\int_{0}^{\infty} dk_{z}\Bigg[ {\frac{d_{e} \, n|Q_{e}|}{\mathcal{E}^{e}_{n}}}
+{\frac{\bar{d}_{e} \, n|Q_{e}|}{\mathcal{E}^{e}_{n}}} \Bigg]
$$
\begin{equation}
-\sum_{f=u}^{d,s}\, {\frac{|Q_{f}|B}{2\pi^{2}}} \sum_{n}3(2-\delta_{n0})\int_{0}^{\infty} dk_{z}\Bigg[ {\frac{d_{f} \, n|Q_{f}|}{\mathcal{E}^{f}_{n}}}
+{\frac{\bar{d}_{f} \, n|Q_{f}|}{\mathcal{E}^{f}_{n}}} \Bigg]
\label{premagnetoalmend}
\end{equation}
$$
p_{\perp}={\frac{3{g_{h}}^{2}}{16{m_{G}}^{2}}}\rho^{2}-{\mathcal{B}}_{QCD} + {\frac{B^{2}}{8\pi}}
+{\frac{|Q_{e}|B^{2}}{2\pi^{2}}} \sum_{n}(2-\delta_{n0})\int_{0}^{\infty} dk_{z}\Bigg[ {\frac{d_{e} \, n|Q_{e}|}{\mathcal{E}^{e}_{n}}}
+{\frac{\bar{d}_{e} \, n|Q_{e}|}{\mathcal{E}^{e}_{n}}} \Bigg]
$$
\begin{equation}
+\sum_{f=u}^{d,s}\, {\frac{|Q_{f}|B^{2}}{2\pi^{2}}} \sum_{n}3(2-\delta_{n0})\int_{0}^{\infty} dk_{z}\Bigg[ {\frac{d_{f} \, n|Q_{f}|}{\mathcal{E}^{f}_{n}}}
+{\frac{\bar{d}_{f} \, n|Q_{f}|}{\mathcal{E}^{f}_{n}}} \Bigg]
\label{perppressuerend}
\end{equation}

\begin{equation}
\rho_{e}={\frac{|Q_{e}|B}{2\pi^{2}}} \sum_{n}(2-\delta_{n0})\int_{0}^{\infty} dk_{z}
\Big(d_{e}-\bar{d}_{e} \Big)
\label{rhoelessfromomegaend}
\end{equation}

\begin{equation}
\rho=\sum_{f=u}^{d,s}\,{\frac{|Q_{f}|B}{2\pi^{2}}} \sum_{n}3(2-\delta_{n0})\int_{0}^{\infty} dk_{z}
\Big(d_{f}-\bar{d}_{f} \Big)
\label{rhoquarksfromomegaend}
\end{equation}
$$
s=-{\frac{|Q_{e}|B}{2\pi^{2}}} \sum_{n}(2-\delta_{n0})\int_{0}^{\infty} dk_{z} \Bigg\{
d_{e} \,  ln(d_{e}) +(1-d_{e})\,ln(1-d_{e})
$$
$$
+\bar{d}_{e} \,  ln(\bar{d}_{e}) +(1-\bar{d}_{e})\,ln(1-\bar{d}_{e})\Bigg\}
$$
$$
-\sum_{f=u}^{d,s} {\frac{|Q_{f}|B}{2\pi^{2}}} \sum_{n}3(2-\delta_{n0})\int_{0}^{\infty} dk_{z} \Bigg\{d_{f} ln(d_{f}) +(1-d_{f})\,ln(1-d_{f})
$$
\begin{equation}
+\bar{d}_{f} \,  ln(\bar{d}_{f}) +(1-\bar{d}_{f})\,ln(1-\bar{d}_{f}) \Bigg\}
\label{sfinalend}
\end{equation}
$$
\varepsilon={\frac{3{g_{h}}^{2}}{16{m_{G}}^{2}}}\rho^{2}+{\mathcal{B}}_{QCD} + {\frac{B^{2}}{8\pi}}
+\,{\frac{|Q_{e}|B}{2\pi^{2}}} \sum_{n}(2-\delta_{n0})\int_{0}^{\infty} dk_{z}\,\mathcal{E}^{e}_{n}(d_{e}+
\bar{d}_{e})
$$
\begin{equation}
+\sum_{f=u}^{d,s} \,{\frac{|Q_{f}|B}{2\pi^{2}}} \sum_{n}3(2-\delta_{n0})\int_{0}^{\infty} dk_{z}\,\mathcal{E}^{f}_{n}(d_{f}+
\bar{d}_{f})
\label{endtotalalmosthafim}
\end{equation}
In the zero temperature limit \cite{dpm3,furn,mags}, applied to astrophysics, we have the distributions (\ref{dists})
given by:
\begin{equation}
d_{i}=\Theta(\nu_{i}-\mathcal{E}^{i}_{n}) \hspace{1.8cm} \textrm{and} \hspace{1.8cm}  \bar{d}_{i}=0
\label{limittempzerodists}
\end{equation}
and also \cite{furn}:
\begin{equation}
\lim_{T \to 0} \,\, T\,ln\Big(1-d_{i}\Big) = (\mathcal{E}^{i}_{n}-\nu_{i})
\hspace{0.8cm} \textrm{and} \hspace{0.8cm}
\lim_{T \to 0} \,\, T\,ln\Big(1-\bar{d}_{i}\Big) =0
\label{limittempzerorels}
\end{equation}

\begin{acknowledgments}

\vskip0.5cm
This work was partially supported by the Brazilian funding agencies CAPES, CNPq and FAPESP.  We thank D\'ebora P. Menezes and
Daryel Manreza Paret for instructive discussions.

\end{acknowledgments}

\end{document}